\documentclass{emulateapj}
\usepackage{apjfonts}
\usepackage{natbib}

\def\plottwo#1#2{\centering \leavevmode
    \epsfxsize=1.0\columnwidth \epsfbox{#1} \hfil
    \epsfxsize=1.0\columnwidth \epsfbox{#2}}

\def\plotthree#1#2#3{\centering \leavevmode
    \epsfxsize=0.67\columnwidth \epsfbox{#1} \hfil
    \epsfxsize=0.67\columnwidth \epsfbox{#2} \hfil
    \epsfxsize=0.67\columnwidth \epsfbox{#3}}



\long\def\comment#1{}

\def\W2{{\cal W}}

\def\be{\begin{equation}}
\def\ee{\end{equation}}
\def\bea{\begin{eqnarray}}
\def\eea{\end{eqnarray}}

\def\Mpc{\,{\rm Mpc}}

\def\cmm2{{\,\rm cm^{-2}}}
\def\cm2{{\,{\rm cm}^2}}
\def\cmm3{{\,{\rm cm}^{-3}}}
\def\gcmm3{{\,{\rm g\,cm^{-3}}}}

\def\fun#1#2{\lower3.6pt\vbox{\baselineskip0pt\lineskip.9pt
  \ialign{$\mathsurround=0pt#1\hfil##\hfil$\crcr#2\crcr\sim\crcr}}}

\def\planck{{\it Planck}}

\def\C{{\cal C}}

\lefthead{CHU \& KNOX}
\righthead{}

\begin{document}
\bibliographystyle{apj}
\submitted{To be submitted to ApJ}
\title{
Testing cosmological models and understanding cosmological parameter 
determinations with metaparameters
}
\author{Mike Chu}
\affil{Department of Physics,
University of California, Davis, CA 95616, USA,
email: mchu@bubba.ucdavis.edu}
\and
\author{Lloyd Knox}
\affil{Department of Physics,
University of California, Davis, CA 95616, USA,
email: lknox@ucdavis.edu}

\begin{abstract}
Cosmological parameters affect observables in physically distinct
ways.  For example, the baryon density, $\omega_b$, affects the
ionization history
and also the pressure of the pre-recombination
fluid.  To investigate the relative importance of different physical effects 
to the determination of $\omega_b$, and to test the cosmological model, 
we artificially split $\omega_b$ into two `metaparameters':
$\omega_{be}$ which controls the ionization history and 
$\omega_{bp}$ which plays the role of $\omega_b$ for
everything else.
In our demonstration of the technique we find
$\omega_b = .0229 \pm .0012$ (with no parameter splitting),
$\omega_{bp} = .0238 \pm .0021$, $\omega_{be}= .0150 \pm .0034$ and
$\omega_{bp}-\omega_{be} = .0088 \pm .0039$. 
\end{abstract}

\keywords{cosmology: theory -- cosmology: observation} 

\section{Introduction}

As predicted \citep{spergel95,knox95,jungman96a}, observations of the
cosmic microwave background (CMB) anisotropies
(e.g. \citet{kuo04,bennett03,readhead04}) have provided very tight
constraints on cosmological parameters
(e.g. \citet{spergel03,goldstein03,rebolo04}).  These constraints are possible
because the statistical properties are sufficiently rich and, given a
model, can be calculated with very high accuracy (e.g. \citet{hu02e}).

One must bear in mind though that these determinations are 
highly indirect and model-dependent.  It is therefore useful to have tools for
testing the model, and for gaining better understanding of the
particular physical processes important for a given constraint.
Toward these ends, we explore use of cosmological `metaparameters'.

A given parameter, $x$, may lead to observational consequences through
more than one distinct physical effect.  Such a parameter can be split
into more than one metaparameter, $x_a$, $x_b$, $x_c$, ... each of
which controls a different physical effect.  This approach to data
analysis has been developed independently and applied recently by
\citet{zhang03} who call it `parameter splitting.'  Their motivation
was to marginalize over physical effects that could not be calculated
with sufficient accuracy.  One can also use the split into
metaparameters to test the model (by checking if $x_a = x_b$ to within
errors), and to understand where the constraints are coming from (by
comparing $\sigma(x_a)$ to $\sigma(x_b)$).

In this paper we explore one example of a split into cosmological
metaparameters.  In particular, we split the baryon density into two
parameters: one that controls the ionization history, $\omega_{be}$,
and one that controls the inertia of the pre-recombination
baryon-photon fluid, $\omega_{bp}$.  In section 2 we describe the
dependence of the angular power spectrum on these two variables.  In
section 3 we briefly describe our calculations.  In section 4 we 
show and discuss our results.  Finally, in section 5 we conclude.  

\section{Dependence of $C_l$ on $\omega_{be}$ and $\omega_{bp}$}

The baryon density affects the evolution of CMB temperature anisotropies
in two distinct ways.  First, in the pre-recombination plasma, the higher
the baryon density, the higher the inertia of the fluid due to the
baryon's mass.  Second, the ratio of baryons to photons determines the
history of the number density of free electrons (prior to reionization).  

These two different physical effects, one related to the baryon's mass and
one related to the large Thomson cross section of the electrons that
charge-balance the baryons, lead to different observational
consequences.  They therefore lead to two different
ways to determine the baryon density.  

We can study these effects separately by splitting $\omega_b \equiv
\Omega_b h^2$ into what we will call $\omega_{be}$ and $\omega_{bp}$
where the $e$ indicates electron and the $p$ stands for pressure.  
Operationally, we calculate the mean number density of electrons, $n_e$, 
assuming $\omega_b = \omega_{be}$, using the recombination
routine RECfast \citep{seager99}.  
Then we calculate $C_l$ using CMBfast \citep{seljak96} with
$\omega_b = \omega_{bp}$ and the $n_e(z)$ from RECfast.

\subsection{Dependence on $\omega_{bp}$}

\begin{figure}[!ht]
\centerline{\scalebox{.4}{\includegraphics{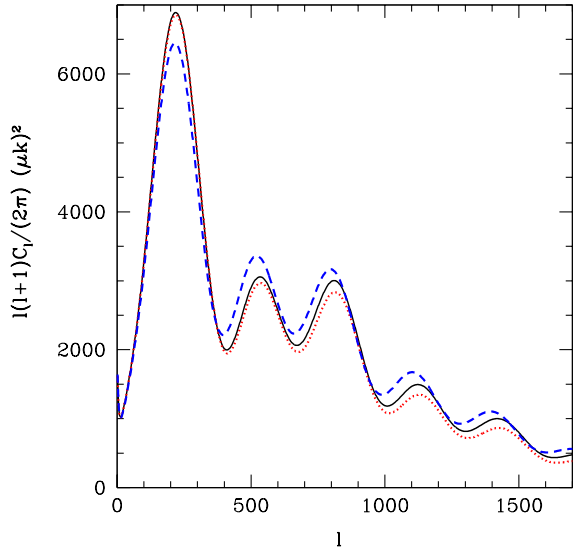}}}
\caption[]{\label{fig:cl}Dependence of $C_l$ on the metaparameters.
Solid curve is $C_l$ for the fiducial model with $\omega_{bp} = \omega_{be} =
0.023$.  Dashed (dotted) curve is $C_l$ for fiducial model parameters
except for $\omega_{bp} = 0.018$ ($\omega_{be} = 0.018$).  The fiducial
model is a typical sample from the MCMC chain of our six-parameter model
space constrained by the WMAPext dataset. 
}
\end{figure}

The response of $C_l$ to the two baryon densities can be seen in
Fig.~\ref{fig:cl}.  First we concentrate on the multiple effects of varying
$\omega_{bp}$.  Decreasing $\omega_{bp}$ leads to three different physical
effects:  sound speed increase, acoustic oscillation zero-point shift, and
reduction in `baryon drag'.  We briefly review the effects here though
they are all discussed at length in the review by \citet{hu02e}.

Adding non-relativistic baryons decreases the sound speed of the fluid,
given by 
\be
c_s = {1 \over \sqrt{3(1+R)}}
\ee
where $R \equiv 3\rho_b/\rho_\gamma$.  Reducing $\omega_{bp}$ decreases $R$
and therefore increases the sound speed, causing the oscillation pattern
to shift slightly to lower $l$.  

In the absence of baryons, if the gravitational potential $\Psi$ is
constant then the effective temperature, $\Theta_0 + \Psi$, oscillates
about zero; that is, the competing effects of pressure and gravity
cancel when $\Theta_0 = -\Psi$.  Adding baryons reduces the pressure,
meaning the baryons must collapse further into a potential well before
gravity and pressure balance.  The zero point shifts to $\Theta_0 =
-(1+R)\Psi$ and therefore $\Theta_0 + \Psi$ oscillates about $-R\Psi$.
The offset enhances odd peaks (for which the effective temperature is
positive in potential hills and therefore $-R\Psi$ is a boost) and
suppresses even peaks (for which the effective temperature is negative
in potential wells).  Decreasing $\omega_{bp}$ therefore suppresses
odd peaks and enhances even peaks.  

In the above paragraph we assumed a constant $\Psi$ which is a good
approximation for the matter-dominated era, but not for radiation
domination.  For modes that enter during radiation domination,
pressure resists transport of material into the potential well,
causing the potential to decay as expansion dilutes the over-density.
Thus for modes that entered early during radiation domination, by
the time of last-scattering the gravitational potential is insignificant
and the oscillation is about $\Theta_0 = 0$; there is no offset to
the oscillations and therefore no modulation of the even and odd peak heights.

At smaller scales, the dominant effect of a reduction of $\omega_{bp}$ is
a reduction in `baryon drag'.  The baryon drag effect can be thought
of as due to the increasing value of $R$ over time.  
As $R$ increases in time, the sound speed decreases, so the oscillation
frequency decreases.  Since energy/frequency of an oscillator is an
adiabatic invariant, this decrease in frequency is matched by a decrease
in oscillation amplitude.  Since $\dot R \propto R$, decreasing $\omega_{bp}$
decreases the amount of baryon drag, and the power is enhanced.  Note
that the baryon drag effect is is distinct from the photon diffusion
we discuss in the next subsection.  This distinction is clear from the
fact that there is baryon drag even in the tight coupling limit.

The scale of matter radiation equality projected to today comes out at
$l_{\rm eq} \simeq 170$ \citep{knox01b} and therefore the first
peak entered slightly before matter-radiation equality.  Both effects
(zero-point offset and baryon drag) are important for the first peak
and partially cancel each other out.  For the second peak, the effects
add so that decreasing $\omega_{bp}$ enhances power.  For the third
peak, having entered earlier during radiation domination, the offset
effect is sub-dominant to baryon drag so power is enhanced.

\subsection{Dependence on $\omega_{be}$}

Even prior to recombination, the mean free path due to Thomson scattering
is non-zero.  Photons thus diffuse, damping the power spectrum on
small scales.  The comoving damping length is given by the distance the
photons can random walk by the time of last scattering. 
Decreasing $\omega_{be}$ decreases the number density of baryons,
$n_b$, and therefore, for fixed ionization fraction, the number 
of free electrons, $n_e = x_e n_b$.  The resulting increase in mean
free path increases the damping length.  

The effect of varying $\omega_{be}$ on the damping length is
complicated by the fact that decreasing $\omega_{be}$ increases the
number of photons per baryon and therefore alters $x_e(z)$.  More
photons per baryon delays recombination.  This means that at a given
time more atoms are ionized (tending to decrease the damping scale),
and it also means a delay in recombination giving more time for the
random walking (tending to increase the damping scale).  \citet{hu97}
found that the net result is that the damping length scale is roughly
proportional to $\omega_b^{-1/4}$.  Reducing $\omega_{be}$ therefore
increases the damping length, leading to the increased suppression for
the dotted curve seen in Fig.~\ref{fig:cl}.  

The damping length projected from the last-scattering surface to here,
a comoving distance $D$ away, gives rise to an angular scale,
$\pi/l_D$.  Below we use $l_D$ as a function of cosmological
parameters as written in \citet{hu01a} with the difference that we
numerically evaluate the redshift of last-scattering, $z_*$, as the
peak of the visibility function, and replace $\omega_b$ with
$\omega_{be}$.

\section{Calculating the Likelihood}

What we want to know is, given the data and any other assumptions we make
about the world, what is the probability distribution of the parameters?
This posterior probability distribution can be calculated by use of 
Bayes' theorem which states:
\be
P({\vec \theta}|d) \propto P(d|{\vec \theta}) P_{\rm prior}({\vec \theta})\,.
\ee
where $d$ refers to data and the proportionality constant is chosen to
ensure  $\int P({\vec \theta}|d) d {\vec \theta}=1$.  
With a uniform prior this simply reduces to 
$P({\vec \theta}|d) \propto P(d|{\vec \theta})$.  
This probability of the data given the parameters is, when thought of
as a function of the parameters, called the likelihood, 
${\cal L}({\vec \theta})$. 

Often we are interested in the posterior probability distribution
for one or two parameters alone.  This marginalized posterior
is given by integrating over the other parameters.  For example:
\be
P(\theta_1,\theta_2|d)= \int \Pi_{i=3}^n d\theta_i {\cal L}(\vec \theta)
P_{\rm prior}(\vec \theta)
\ee
where $n$ is the number of parameters.  We use the
prior to incorporate non--CMB information such as that the redshift of
reionization must be greater than 6.0 \citep{becker01}.

In the following  subsections we discuss first how we evaluate the
likelihood function at a single point and then how we evaluate it over
a large parameter space and produce marginalized posterior
distributions. 

\subsection{Likelihood evaluation}

The first step to likelihood evaluation is to calculate
the angular power spectrum, $C_l$, for the cosmological model.
To do this we use CMBfast.  Despite its speed advantages
we do not use the Davis Anisotropy Shortcut (DASh; \citet{kaplinghat02}) 
since the split $\omega_b$ modifications were easier with
CMBfast than with DASh.

Once $C_l$ is calculated we evaluate the likelihood given the WMAP
data with the subroutine available at the LAMBDA\footnote{ The Legacy
Archive for Microwave Background Data Analysis can be found at
http://lambda.gsfc.nasa.gov/} data archive.  For CBI and ACBAR we use
the offset log-normal approximation of the likelihood \citep{bond00}.
The likelihood given all these data together (referred to as the
WMAPext dataset in \citet{spergel03}) is given by the product of the
individual likelihoods.

We do not use the most recent release of CBI data \citep{readhead04},
nor the VSA data \citep{dickinson04}.  We have used the older release
\citep{pearson03} for ease of comparison with results in
\citet{spergel03}.  The new CBI data \citep{readhead04} and the VSA
data are consistent with the old CBI data, WMAP and ACBAR
\citep{dickinson04,rebolo04}.

\subsection{Exploring the Parameter Space}

We explore several parameter spaces.  What we refer
to as the six-parameter model has six cosmological parameters
(the baryon density, the cold dark matter density, 
the scalar primordial power spectrum amplitude at $k=0.05 \Mpc^{-1}$, $n_S$, 
the redshift of reionization and
$\Omega_\Lambda$) and a calibration parameter for each of CBI and
ACBAR.  The split model is the same except the baryon density, $\omega_b$, 
is replaced with $\omega_{be}$ and $\omega_{bp}$.  

We explore the parameter space by producing a Monte Carlo Markov Chain
(MCMC) via the Metropolis--Hastings algorithm as described in
\citet{christensen01}.  Our procedure is the same as in \citet{chu03}
except for some changes to the adaptive phase of the sampling, during
which the generating function is determined.  Our WMAPext chain
for the six-parameter model space has 80,000 samples and the WMAPext
chain for the seven-parameter split model space has 130,000 samples.  

\section{Results}

We now examine how the likelihood function changes as the
parameter space is expanded to non-zero $\omega_{bp}-\omega_{be}$.

\subsection{Constraints on $\omega_{be}$ and $\omega_{bp}$}

In Fig.~2
we see the resulting constraints on
$\omega_b$ (assuming no splitting), $\omega_{be}$ and $\omega_{bp}$.
Note that, without splitting, we find $\omega_b = 0.0229 \pm 0.0012$ which
reproduces the result from \citet{spergel03} of $0.023 \pm 0.001$ for
the same model space and dataset.  

For the unsplit parameter there is a tension between increasing the
damping length (which occurs by lowering $\omega_b$) and increasing
inertial effects (which occurs by raising $\omega_b$).  
This tension only becomes evident once we split the parameter and see
that $\omega_{be}$ drifts downward and $\omega_{bp}$ drifts slightly
upward.

\begin{figure*}
\label{fig:result}
\plottwo{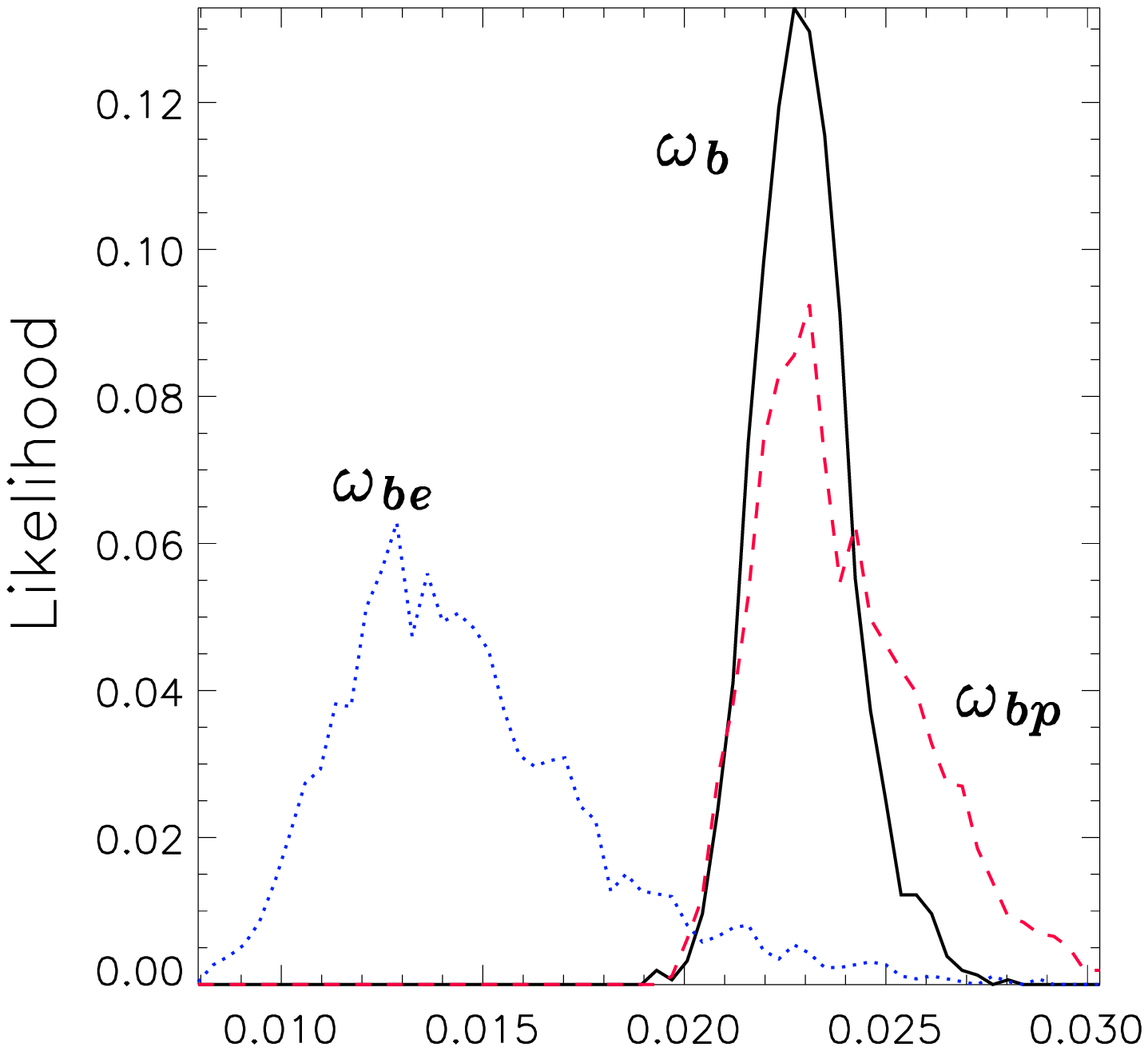}{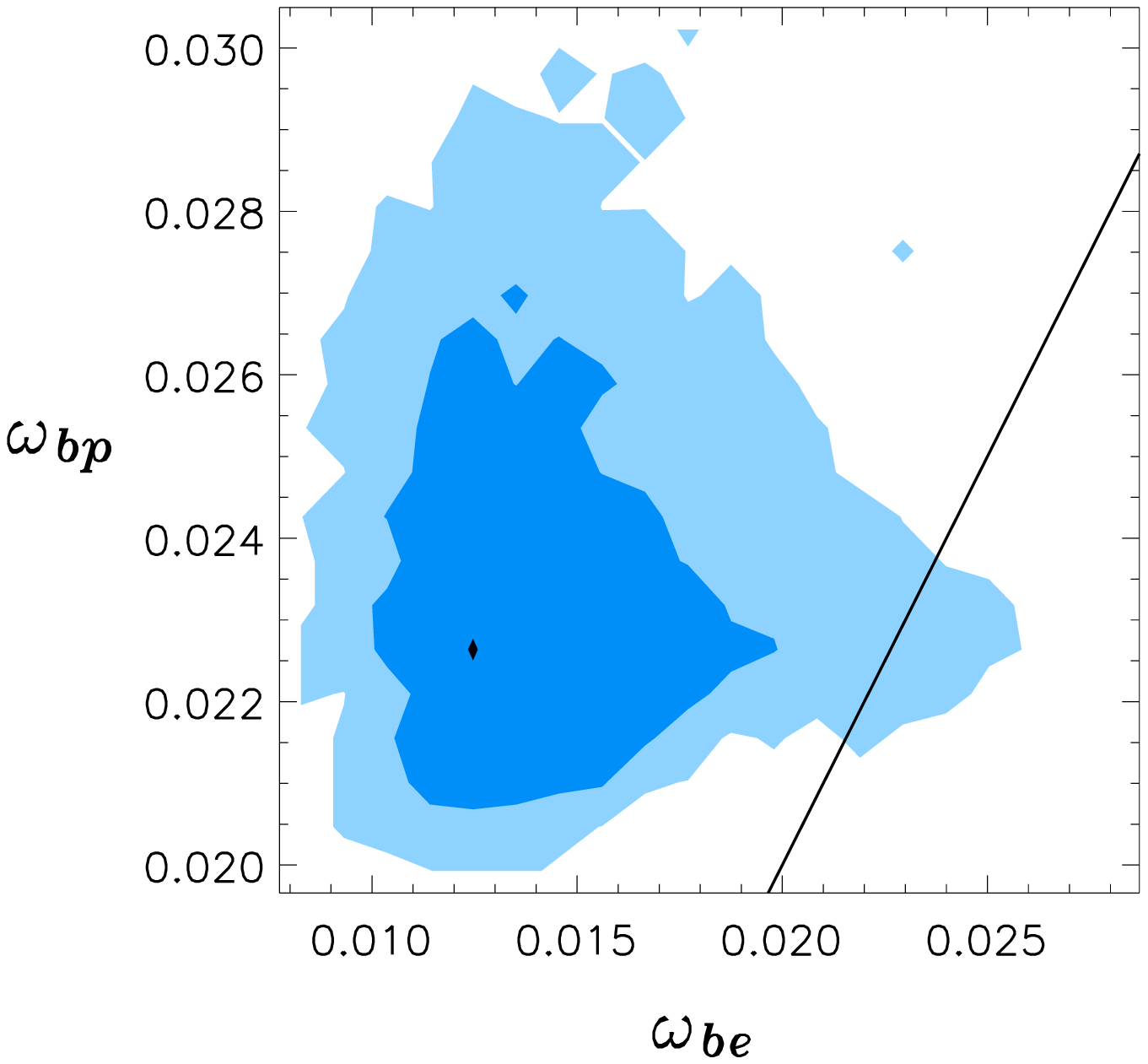}
\caption{Left panel: One-dimensional marginalized
likelihoods of $\omega_b$ (assuming no parameter splitting), $\omega_{be}$
and $\omega_{bp}$.  Right panel:  Two-dimensional marginalized likelihood
of $\omega_{bp}$ vs. $\omega_{be}$.  Dark (light) 
shaded region is where the likelihood
is down by less than a factor of $\exp(-2.3/2)$ ($\exp(-6.17/2)$)
from its maximum value.  The line is $\omega_{bp}=\omega_{be}$.
}
\end{figure*}

\begin{figure}[!ht]
\centerline{\scalebox{.5}{\includegraphics{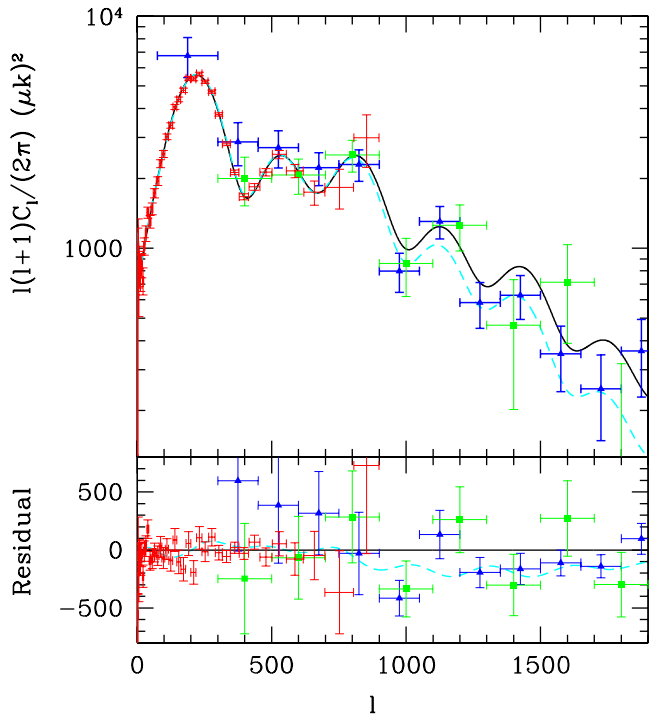}}}
\caption[]{\label{fig:bandpowers}Top panel: CMB constraints on the
power spectrum, the fiducial six-parameter model, and an improved fit
that comes from allowing $\omega_{bp} \ne \omega_{be}$. Bottom panel:
Same as above, but with the fiducial six-parameter model subtracted.}
\end{figure}

In Fig.~\ref{fig:bandpowers} we see why the data tend to favor
lower $\omega_{be}$:  it improves the fit in the damping tail
region.  Model fits with the six-parameter model lead to
excess model power at higher $l$.  Allowing $\omega_{be}$ to
vary independently from $\omega_{bp}$ gives the necessary freedom
to eliminate this excess and improve the over all fit.  The
tendency for $\omega_{bp}$ to drift upward, seen in Fig.~2, 
has indirect causes as we explain in the next subsection.

The excess small-scale power of the six-parameter model fits
is also what causes the slight preference of the data for
$dn_S/d\ln k < 0$.  \citet{spergel03} find for WMAPext that
$dn_S/d\ln k = -0.055 \pm 0.038$.  This extension by splitting
$\omega_b$ actually leads to somewhat better agreement with the
data than the extension to $dn_S/d\ln k \ne 0$.  We find the 
maximum likelihood improves by a factor of 8.2 with the
extension to split $\omega_b$ and 2.8 with the extension to
$dn_S/d\ln k \ne 0$.

In the right panel of Fig.~2 
we see the constraints in the $\omega_{bp}$-$\omega_{be}$ plane.
The straight line is the physical subspace, $\omega_{bp}=\omega_{be}$.  
How significant is the deviation from the physical subspace?
We find $\omega_{bp}-\omega_{be} = 0.0088 \pm 0.0039$, a 2.3$\sigma$
difference from zero.  The difference from zero could simply be
a statistical fluctuation, it could be caused by systematic error
in one or more of the experiments, or it could be that 
the six-parameter model does not adequately describe reality.
The discrepancy is not strong enough to rule out the first option.

\subsection{Constraints on other parameters}

Splitting $\omega_b$ alters the constraints on other parameters too.  
To gain a better understanding of the model dependence we now
examine how constraints on $\tau$, $n_S$ and $\omega_m$ change.

In the six-parameter model there is a tension between the value
of $\tau$ that best fits the $C_l^{TT}$ data ($\tau = 0$) and the larger
value which best fits the $C_l^{TE}$ data.  Allowing
extra freedom in the damping tail, by splitting $\omega_b$, 
makes it possible for a higher
$\tau$ to be consistent with the $C_l^{TT}$ data as can be seen
in Fig.~4. 
Increasing $n_S$
can compensate for increased $\tau$, but too large an $n_S$ leads
to too much power on small scales.  Extra damping from decreased
$\omega_{be}$ compensates for this excess power.  Further, increased
$n_S$ increases the ratio of the 2nd peak height to first peak
height, and therefore $\omega_{bp}$ increases to compensate.

\begin{figure*}
\label{fig:otherparams}
\plotthree{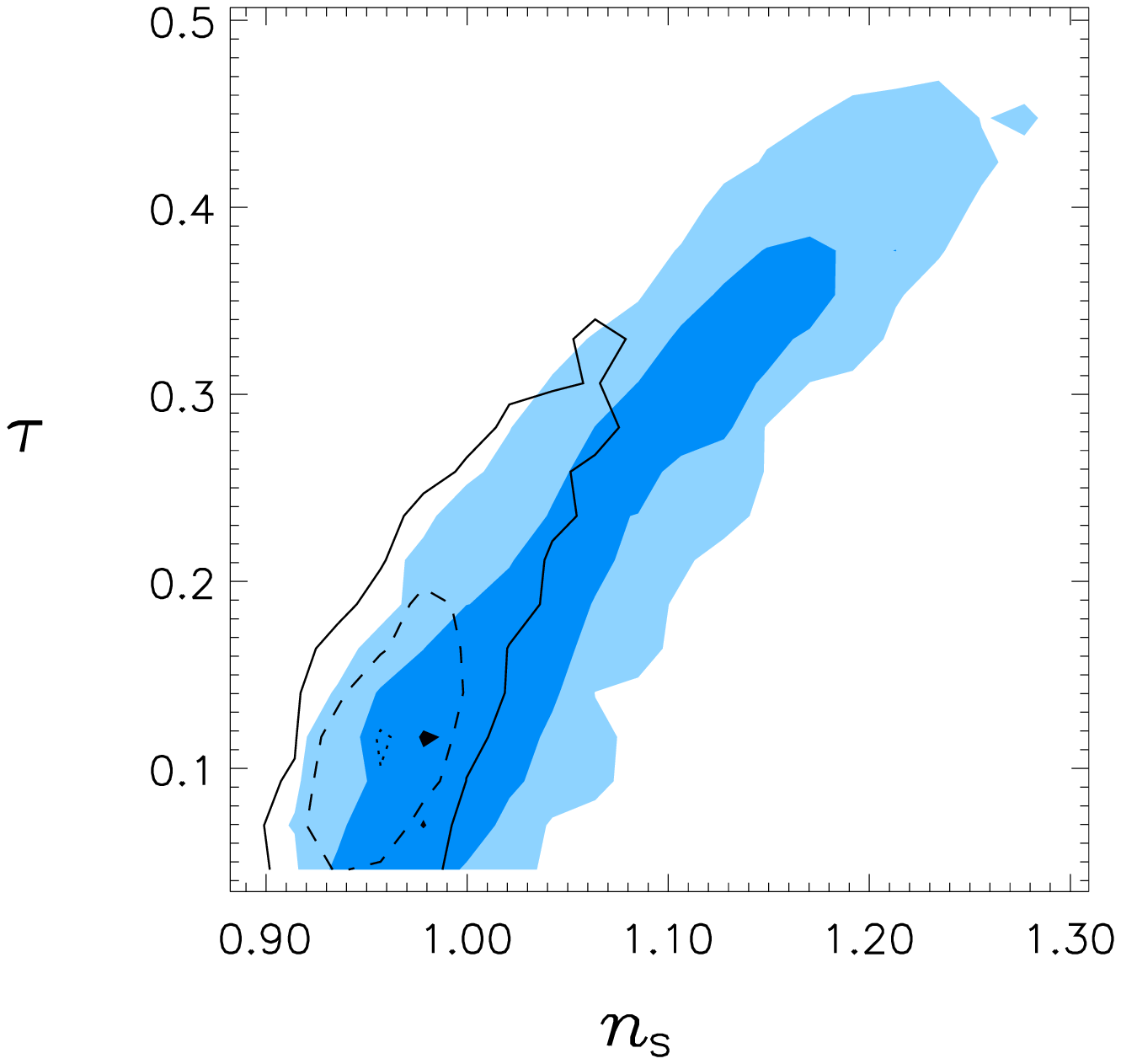}{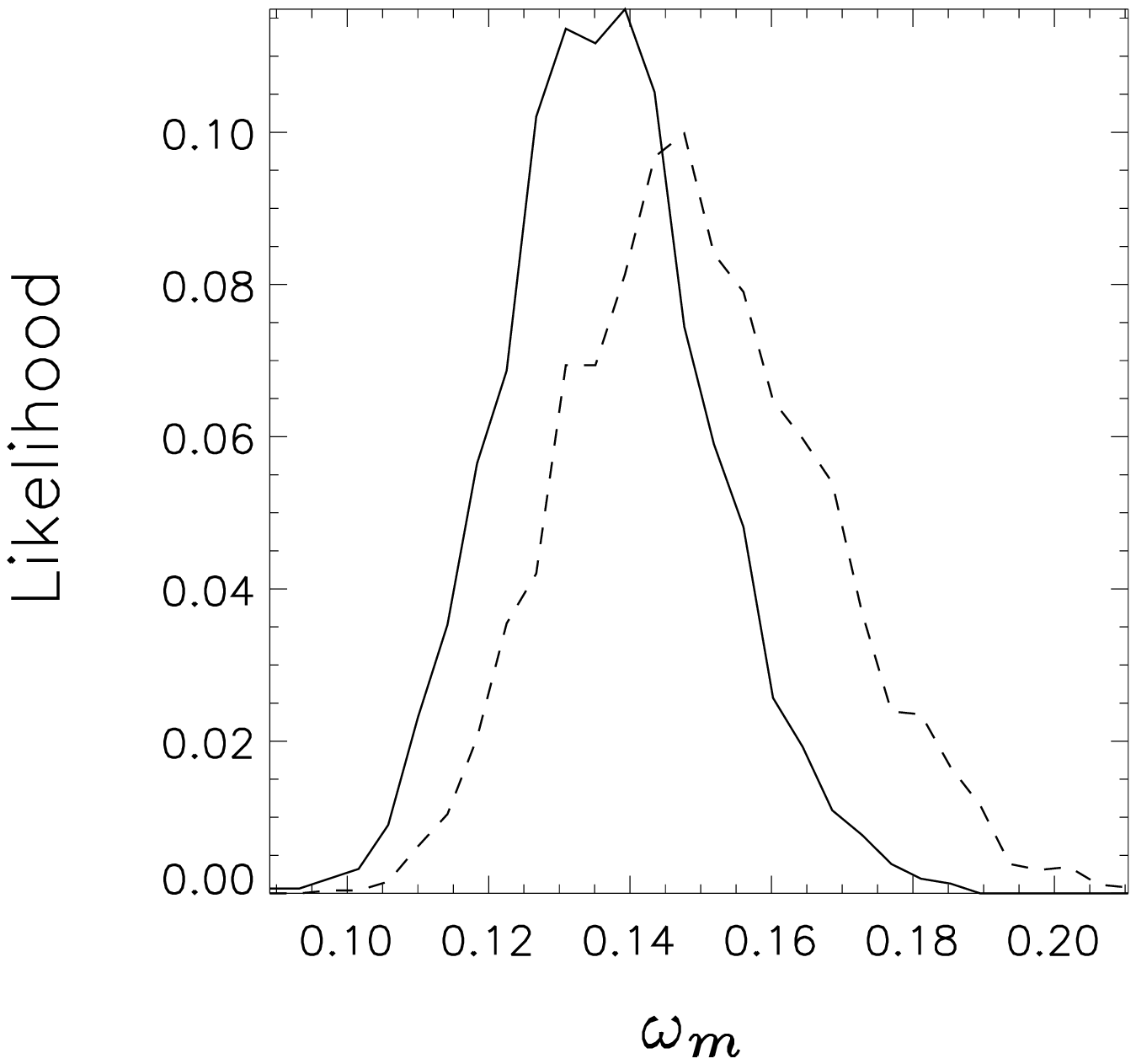}{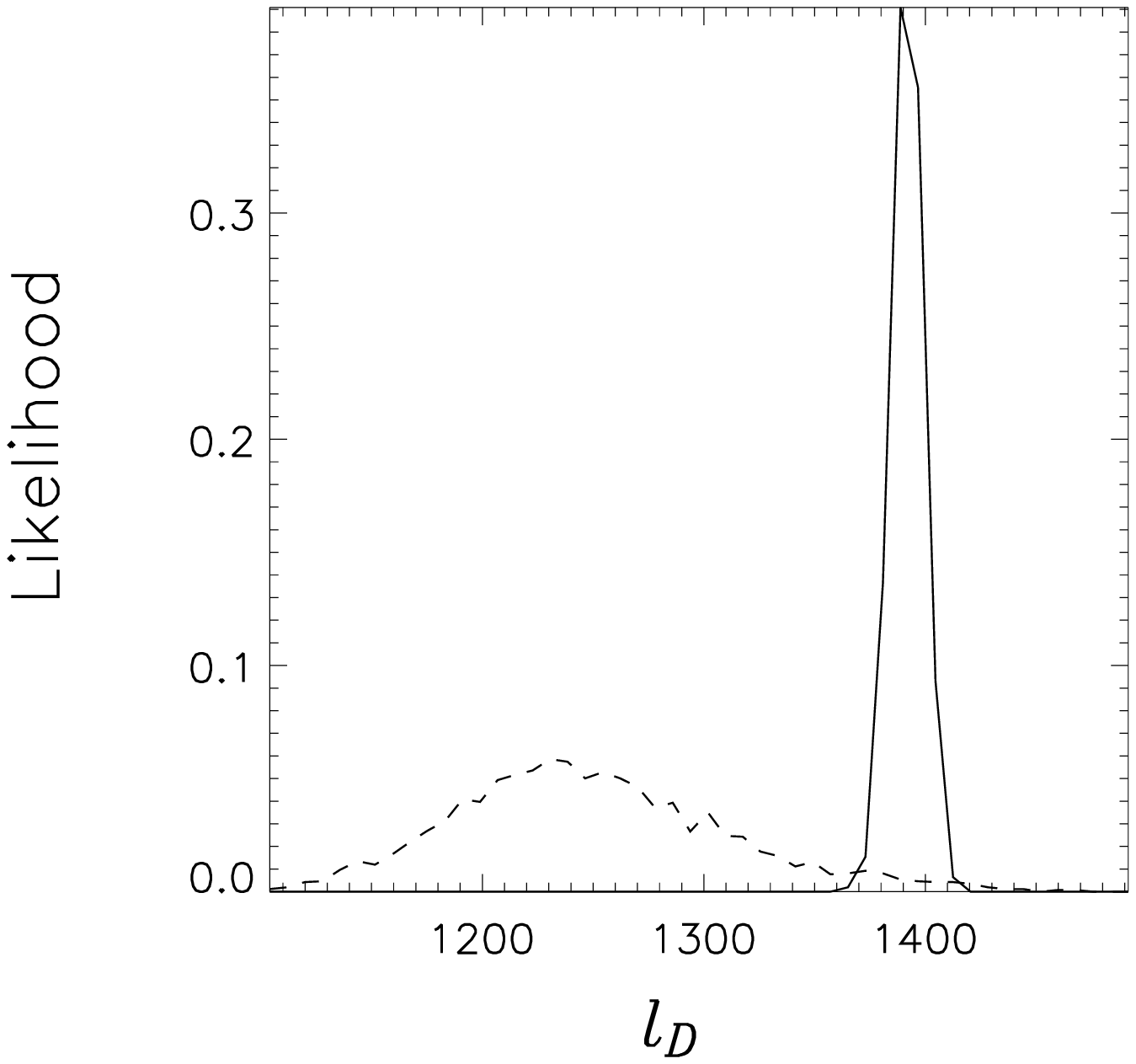} 
\caption{Two- and one-dimensional marginalized likelihoods. 
Left panel:  The dashed (solid) line is where the likelihood
of $n_s$ and $\tau$ given the WMAPext dataset assuming the six-parameter
model is down by a factor of $\exp(-2.3/2)$ ($\exp(-6.17/2)$)
from its maximum value.  Shading indicates the same regions,
but for the split model space.  Middle and right panels:  
One-dimensional marginalized likelihoods of $\omega_m$ and $l_D$ given
the WMAPext dataset.  Solid lines are for the six-parameter model space and
dashed lines are for the split model space.
}
\end{figure*}


We also see in Fig.~\ref{fig:otherparams} that the likelihood function
of $\omega_m$ has broadened some toward higher values.  \citet{spergel03}
claim that the constraint on $\omega_m$ is largely coming from the
rise to the first peak and how the shape of the spectrum in this region
is influenced by the early integrated Sachs-Wolfe (ISW) effect.  The amount
of fluctuation power from the early ISW effect depends on the ratio
of matter energy density to radiation energy density at last scattering.  We
see from our split model chains that models with higher $z_*$ also 
tend to have higher $\omega_m$, as they must to keep $\rho_m/\rho_z$
at $z_*$ nearly fixed.  

\subsection{Benefits of measuring the damping tail}

Measuring the damping tail region of the spectrum to high precision
will be very valuable. In the context of the six parameter model
it will allow for much tighter constraints on $\omega_b$, $\omega_m$
and $n_S$.  Whereas WMAP will measure out to about $l=500$ with
cosmic variance precision, \planck ~\citep{planck} will 
measure out to about $l=2000$
with cosmic variance precision, reducing the errors on the above
parameters by factors of 6, 5 and 6 respectively 
\citep{eisenstein99}.

Perhaps more importantly, measuring the damping tail will allow
more stringent tests of the model.  The six-parameter
model, calibrated with measurements at larger angular scales,
makes tight predictions for the damping tail region.  For
example, the damping scale $l_D$, is very tightly constrained
with the WMAPext data given the six-parameter model, as can
be seen in Fig.~\ref{fig:otherparams}.
But these tight constraints are not because we have measured the
damping tail well.  They are due to the fact that the parameters
controlling the damping tail region can be determined well
at larger angular scales.  

With the split model we weaken this connection between the acoustic
region and the damping region.  As a result, the constraints on $l_D$
weaken considerably as seen in Fig.~\ref{fig:otherparams} .  Future
high precision measurements of the damping tail will be able to
determine $l_D$ with high accuracy.  Then the prediction for $l_D$,
given the six-parameter model and acoustic region data, can be
compared with $l_D$ inferred from the damping region, allowing for a
strong test of the model.  

Current measurements of the damping tail are already playing an
important role in testing the six-parameter model.  ACBAR and CBI do
reduce the allowed region of the six-dimensional parameter space some,
but mostly serve to confirm the WMAP predictions (albeit with slightly
lower power).  Once we allow for the parameter split, ACBAR and CBI
place significant extra constraints on the parameter space.  For
example, constraints on the split $\omega_b$ model from the WMAP data
mean that $\C_{1400} = 805 \pm 169$ (where $\C_l \equiv
l(l+1)C_l/(2\pi)$).  Inclusion of the ACBAR and CBI data change this
to $695 \pm 79$, reducing the uncertainty by a factor of two.  In
contrast, without the $\omega_b$ split, adding ACBAR and CBI only
reduces the $\C_{1400}$ error by 30\%.

\section{Conclusions}

We have introduced the creation of cosmological metaparameters
as a tool for exploring the powerful, complicated, model-dependent 
constraints on cosmological parameters possible with measurements
of CMB anisotropy.  Such an exploration is useful for gaining
a physical understanding of the origin of the constraints and for
testing the consistency of the model.  We see, as expected, $\omega_b$
is mostly constrained via the observable consequences of its effect
on the inertia of the pre-recombination plasma.  Electron-scattering
effects play a sub-dominant role.  Determinations of $\omega_b$ from
the two different effects differ by 2.3$\sigma$; i.e., they are
marginally consistent with each other.

It is amazing that a model with only six parameters can provide such a
good fit to the WMAPext dataset.  With the assumption of this model,
fairly tight constraints are possible on parameters such as $n_S$ and
$\omega_b$.  We see though, as others have seen \citep{tegmark04},
that extending the model space by just one parameter can greatly
broaden the constraints on other parameters.  Interpretation of CMB
constraints on cosmological parameters must be done with care.

In particular we have seen that the modeling of processes that
affect the damping tail region of the spectrum can greatly
loosen bounds on $n_S$ and $\tau$.  Other (physical) parameter extensions
that would affect the damping tail region are running, and
a time-varying fine structure constant \citep{kaplinghat99}.  Measurement
of the damping tail with great precision, such as will be done
by \planck, will dramatically decrease the sensitivity to modeling 
uncertainty and provide stringent consistency tests.

\acknowledgments

We thank M. Kaplinghat \& A. Stebbins for useful conversations and 
M. Kaplinghat for comments on the manuscript. This work was supported by NASA grant NAG5-11098.

\bibliography{/work3/knox/bib/cmb3}

\end{document}